# Hartree-Fock-Bogoliubov Theory of Polarized Fermi Systems

George Bertsch,[1] Jacek Dobaczewski,[2,3] Witold Nazarewicz,[4,5,2] and Junchen Pei[6,4,5]

[1]*Institute for Nuclear Theory and Department of Physics,*
*University of Washington, Seattle, WA 98195-1560*
[2]*Institute of Theoretical Physics, Warsaw University, ul. Hoża 69, 00-681 Warsaw, Poland*
[3]*Department of Physics, P.O. Box 35 (YFL), FI-40014 University of Jyväskylä, Finland*
[4]*Department of Physics & Astronomy, University of Tennessee, Knoxville, Tennessee 37996, USA*
[5]*Physics Division, Oak Ridge National Laboratory,*
*P.O. Box 2008, Oak Ridge, Tennessee 37831, USA*
[6]*Joint Institute for Heavy Ion Research, Oak Ridge National Laboratory, Oak Ridge, TN 37831, USA*
(Dated: October 22, 2018)

Condensed Fermi systems with an odd number of particles can be described by means of polarizing external fields having a time-odd character. We illustrate how this works for Fermi gases and atomic nuclei treated by density functional theory or Hartree-Fock-Bogoliubov (HFB) theory. We discuss the method based on introducing two chemical potentials for different superfluid components, whereby one may change the particle-number parity of the underlying quasiparticle vacuum. Formally, this method is a variant of non-collective cranking, and the procedure is equivalent to the so-called blocking. We present and exemplify relations between the two-chemical-potential method and the cranking approximation for Fermi gases and nuclei.

## I. INTRODUCTION

Due to the experimental realization of strongly interacting atomic Fermi gases, there has been increased theoretical interest in properties of asymmetric two-component superfluid Fermi systems with unusual pairing configurations [1, 2, 3, 4, 5, 6, 7, 8, 9, 10, 11]. Of current interest are the properties of spin-polarized condensates having an unequal number of spin-up and spin-down fermions. One of the condensation possibilities is the "breached-pair" (BP) superfluid state, in which the normal phase coexists with a full pairing of the minority fermions at high spin imbalances. Recently, a gap in a single-particle excitation spectrum of a highly spin-imbalanced sample has been observed experimentally in a $^6$Li atomic condensate [12].

Atomic nuclei can also exhibit interesting pairing properties, including BP superfluidity, although the fraction of the polarization is typically quite small. Examples include:

- Pairing isomers [13, 14, 15] having an appreciably smaller value of the pairing gap than the ground state. In the roots of pairing isomerism is very different single-particle level density in the vicinity of the Fermi surface in coexisting configurations, usually associated with different nuclear shapes. Such states have been found in, e.g., $A{\approx}80$, 100, and 190 mass regions.

- Gapless superconductivity at high spins of atomic nuclei [16, 17]. Here, quasiparticle energies in the rotating frame become negative at certain rotational frequencies; i.e., there is no gap in the quasiparticle spectrum. However, the pairing energy, or the expectation value of the pairing condensate, still remains large.

- Angular-momentum aligned nuclear configurations, such as high-spin isomers, in which pairing correlations may be reduced due to blocking (see Refs. [18, 19, 20] and references quoted therein).

In this paper we will briefly rederive some of the generic results of the Hartree-Fock-Bogoliubov theory (also called the Bogoliubov de-Gennes theory) for polarized quasiparticle states, using constraining fields to reach the states, as was done in Refs. [7, 10] to study the polarized atomic condensates. We shall call this the "two-Fermi level approach" (2FLA). In particular, odd-mass systems require special attention in the HFB theory. In the usual blocking approach, the selected quasiparticle state becomes occupied and this requires a modification of the HFB density matrix. In 2FLA, the lowest quasiparticle orbit becomes occupied by changing the particle-number parity of the vacuum through the external field without modifying any of the vectors explicitly. We show that for spin systems (e.g., Fermi gases) the 2FLA is equivalent to the standard rotational cranking approximation while this is not the case for atomic nuclei, in which polarization is due to the angular momentum alignment. In both cases, however, 2FLA can be viewed as a "vacuum selector" by means of the non-collective cranking.

The paper is organized as follows. Section II briefly discusses the concept of Bogoliubov quasiparticles. Particular attention is paid to the choice of the HFB vacuum, the way the particle-number parity is encoded in the Bogoliubov matrix transformation, and self-consistent signature symmetry of HFB and its relation to time reversal. The HFB extension to the case of two-component systems (2FLA) is outlined in Sec. III, and its relation to non-collective cranking and the blocking procedure is discussed. Section IV shows numerical examples, both from atomic and nuclear physics, based on the HFB approach. Finally, Sec. V contains the main conclusions of this work.

## II. THE QUASIPARTICLE FORMALISM

We begin by recalling basic equations of the quasiparticle formalism, which historically is attributed to Gor'kov, Bogoliubov, and de Gennes. While these equations and definitions are admittedly very well known, there are several aspects of the quasiparticle approach that are seldom discussed; hence, they are worth bringing to the attention of a wider community. We shall discuss these lesser-known aspects in the subsections following the general introduction to HFB.

The HFB wave functions are quasiparticle product states. The quasiparticle annihilation operators $\alpha_\mu$ are defined as linear combinations of particle annihilation and creation operators by the Bogoliubov transformation,

$$\alpha_\mu := \sum_\nu \left( U^*_{\nu\mu} a_\nu + V^*_{\nu\mu} a^+_\nu \right). \tag{1}$$

The matrices $U$ and $V$ satisfy the following canonical conditions:

$$U^+ U + V^+ V = 1, \tag{2a}$$
$$U^+ V^* + V^+ U^* = 0, \tag{2b}$$
$$U U^+ + V^* V^T = 1, \tag{2c}$$
$$U V^+ + V^* U^T = 0. \tag{2d}$$

The HFB vacuum $|\Psi\rangle$ is a zero-quasiparticle state:

$$\alpha_\mu |\Psi\rangle = 0. \tag{3}$$

Complete information about $|\Psi\rangle$ is, in fact, contained in the generalized density matrix $\mathcal{R}$,

$$\mathcal{R} := \langle \Psi | \begin{pmatrix} a^+_\mu a_\nu & , & a_\mu a_\nu \\ a^+_\mu a^+_\nu & , & a_\mu a^+_\nu \end{pmatrix} | \Psi \rangle = \begin{pmatrix} \rho_{\nu\mu} & , & \kappa_{\nu\mu} \\ \kappa^+_{\nu\mu} & , & 1 - \rho^*_{\nu\mu} \end{pmatrix}, \tag{4}$$

which in terms of matrices $U$ and $V$ reads:

$$\mathcal{R} = \begin{pmatrix} V^* V^T & , & V^* U^T \\ U^* V^T & , & U^* U^T \end{pmatrix} = \begin{pmatrix} V^* \\ U^* \end{pmatrix} \begin{pmatrix} V^T & U^T \end{pmatrix}. \tag{5}$$

The variational principle implies that the self-consistent density matrix $\mathcal{R}$ commutes with the quasiparticle Hamiltonian

$$\mathcal{H} := \begin{pmatrix} h - \lambda I & , & \Delta \\ \Delta^+ & , & -h^* + \lambda I \end{pmatrix}, \tag{6}$$

where $h = t + \Gamma$ is the single-particle Hamiltonian and $\Gamma$ and $\Delta$ are particle-hole and particle-particle mean-fields, respectively. The HFB equations can be written in a matrix form:

$$\begin{pmatrix} h - \lambda I & , & \Delta \\ \Delta^+ & , & -h^* + \lambda I \end{pmatrix} \begin{pmatrix} U & , & V^* \\ V & , & U^* \end{pmatrix} = \begin{pmatrix} U & , & V^* \\ V & , & U^* \end{pmatrix} \begin{pmatrix} E & , & 0 \\ 0 & , & -E \end{pmatrix}, \tag{7}$$

where $E$ is a diagonal matrix of quasiparticle energies $E_\mu$. Columns of eigenvectors,

$$\varphi := \begin{pmatrix} V^* \\ U^* \end{pmatrix}, \quad \chi := \begin{pmatrix} U \\ V \end{pmatrix}, \tag{8}$$

are called occupied and empty quasiparticle states, respectively, because they are eigenvectors of the projective matrix $\mathcal{R}$ with eigenvalues 1 and 0.

### A. Choice of occupied quasiparticle states

In practical applications, one often assumes that the quasiparticle energies in matrix $E$ are all positive. However, this is not at all required by the variational principle. Indeed, while eigenvalues of $\mathcal{H}$ must come in pairs of opposite quasiparticle energies $(E_\mu, -E_\mu)$, the theory says nothing on whether a positive or a negative one enters matrix $E$.

This question is evidently related to the mundane problem of the order in which we arrange eigenvectors of $\mathcal{H}$ in Eq. (7). Any such order is allowed, provided one eigenvector of the pair $(E_\mu, -E_\mu)$ is put into the first half of the spectrum (empty quasiparticle states), and the other one is put in the second half of the spectrum (occupied quasiparticle states). However, which one goes where is in principle arbitrary. The signs of quasiparticle energies in $E$ are arbitrary too, and thus variational equations have not one, but many solutions. Naturally, in practical applications, the choice of occupations is motivated by physics as the energy of the system may be different depending on how the arbitrariness of choosing signs of quasiparticle energies is resolved. It is then plausible, and often assumed, that the lowest total energy is obtained by occupying states corresponding to negative quasiparticle energies, and leaving empty those corresponding to positive quasiparticle energies, i.e., by assuming that positive quasiparticle energies are collected in matrix $E$. One should stress at this point, however, that there is no a priori reason why such a choice must *guarantee* obtaining the lowest total energy, and examples to the contrary are available in numerous applications. Thus, which quasiparticle states should be occupied is a matter of a specific physical situation, and not a rule cast in stone.

The problem of choosing occupied quasiparticle states is particularly conspicuous when iterative (self-consistent) methods are used for solving the non-linear eigenproblem (7). The self-consistent procedure can be described as the sequence of the following steps:

1. Find eigenstates of the quasiparticle Hamiltonian (6).

2. Choose the occupied quasiparticle states.

3. Calculate the generalized density matrix (5), i.e., the density matrix ($\rho$) and pairing tensor ($\kappa$).

4. Calculate the particle-hole ($\Gamma$) and particle-particle ($\Delta$) mean fields to determine the Hamiltonian (6) in the next iteration.

5. Return to step 1.

Again, this self-consistent sequence of steps is described in all textbooks; however, the crucial step 2 is seldom ever mentioned. However, in many applications, this step is absolutely essential for obtaining a convergent algorithm. Moreover, choosing negative occupied quasiparticle energies at each iteration may sometimes lead to divergent iterations.

The problem here is in finding ways of identifying quasiparticle states (tags), which are independent of their energies, and which would allow for pinning down the structure of each individual state. Ideally, tags could be provided by quantum numbers of conserved symmetries. For example, spherical symmetry allows each state to be characterized by the standard quantum numbers $nljm$. Then, decisions of occupying quasiparticle states can be made based on lists of quantum numbers.

A particular version of using conserved quantum numbers as tags of quasiparticle states was pioneered in nuclear physics under the name of non-collective rotation, or non-collective cranking approximation [17, 18, 21], and this method is a central theme of our discussion. It consists of using in Eq. (6) the single-particle Routhian $h'$ instead of the single-particle Hamiltonian $h$, where

$$h' := h - \lambda_S S, \qquad (9)$$

and $S$ is the operator of a conserved symmetry. This may look like an attempt of introducing a Lagrange multiplier to constraining the average value of $S$, but in fact it is not. Indeed, since quasiparticle states are eigenstates of $S$, the quasiparticle energies are simply reordered as a function of $\lambda_S$ without any modification of the structure of quasiparticle states or total energy. Using Routhian $h'$ may thus facilitate selecting proper quasiparticle states into the set of occupied ones, depending on the physical situation corresponding to the chosen symmetry $S$. In short, non-collective cranking can be viewed as a configuration selector.

The method fails when quasiparticle states are not eigenstates of symmetry operators, which can happen in unrestricted calculations. The practical solution consists of calculating overlaps of quasiparticle states with a fixed set of wave-functions and establishing in this way tags that are related to the closest similarities of structure between both sets.

Only when a set of properly tagged quasiparticle states is occupied at step 2 of each iteration can one have a good chance of obtaining a converged solution. At the end of the day, one can check whether such a solution corresponds to the negative quasiparticle states being occupied or not.

### B. States with even and odd number of fermions

The ambiguity of choosing the occupied quasiparticle orbits is particularly important for odd-particle systems. Equations (1)–(8) look entirely the same irrespective of whether the state $|\Psi\rangle$ represents even- or odd-particle systems. While the particle number $N$ is not conserved by the product state $|\Psi\rangle$, i.e., this state is a linear superposition of components having different particle numbers, the parity of the particle number, $\pi_N = (-1)^N$, is conserved. That is, the decomposition of $|\Psi\rangle$ may contain either even- or odd-particle number components, but never both.

Of course, information on whether the particular vacuum $|\Psi\rangle$ is $\pi_N$-even or odd is contained in matrices $U$ and $V$ of the Bogoliubov transformation (1), i.e., it must depend on choices made for the occupied quasiparticle states. Specific choices of occupied quasiparticle states must be made in order to obtain even ($\pi_N = 1$) or odd ($\pi_N = -1$) states.

In which way is the information on $\pi_N$ encoded in matrices $U$ and $V$? In the simplest situation, the set of occupied quasiparticle states are such that $U$ is non-singular, $\det(U) \neq 0$. Then we can determine the matrix $Z = VU^{-1}$, and express the vacuum $|\Psi\rangle$ through the Thouless theorem [17]:

$$|\Psi\rangle = |\Phi\rangle_{\text{even}} = \mathcal{N} \exp\left(-\tfrac{1}{2} \sum_{\mu\nu} Z^+_{\mu\nu} a^+_\mu a^+_\nu\right) |0\rangle, \qquad (10)$$

where the normalization factor is $\mathcal{N} = \det^{1/2}(U^+)$. From this expression we see that $|\Psi\rangle$ has even number parity, $\pi_N = 1$.

Let us now make another choice of selecting the occupied states: we replace *one* ($\mu$th) column of the matrix of occupied states $\varphi$ by the same ($\mu$th) column of the matrix of empty states $\chi$ (see Eq. (8)), i.e.,

$$U^{(\mu)}_{\nu'\nu} = \begin{cases} U_{\nu'\nu} & \text{for } \nu \neq \mu, \\ V^*_{\nu'\nu} & \text{for } \nu = \mu, \end{cases} \quad (11a)$$

$$V^{(\mu)}_{\nu'\nu} = \begin{cases} V_{\nu'\nu} & \text{for } \nu \neq \mu, \\ U^*_{\nu'\nu} & \text{for } \nu = \mu, \end{cases} \quad (11b)$$

In this way, we replace the quasiparticle energy of the occupied state $E_\mu$ by $-E_\mu$. We do not imply here that $E_\mu$ must have been positive, so we have replaced a positive quasiparticle energy by a negative one – we could have just replaced the negative one by the positive one. In fact, as discussed earlier, there is no such rule that occupations leading to the original non-singular matrix $U$ must correspond to all quasiparticle energies being positive.

Now it is a matter of simple algebra to see that for such a choice of occupied quasiparticle states, the vacuum state reads

$$|\Phi\rangle^{(\mu)}_{\text{odd}} = \mathcal{N}\alpha^+_\mu \exp\left(-\tfrac{1}{2}\sum_{\mu\nu} Z^+_{\mu\nu} a^+_\mu a^+_\nu\right)|0\rangle, \quad (12)$$

and is a manifestly odd state, $\pi_N = -1$, which we call a one-quasiparticle excitation of $|\Phi\rangle_{\text{even}}$.

Note that in the odd state (12), the matrix $Z$ and normalization constant $\mathcal{N}$ are defined through the original matrices $U$ and $V$ of the even state (10), and not through those after the column replacement as in Eq. (11). Indeed, it is easy to see that the Thouless theorem does not work for the one-quasiparticle states because matrices $U^{(\mu)}_{\nu'\nu}$ are singular. This is obvious from Eq. (2d), whereby each column of matrix $V^*$ is a linear combination of columns of matrix $U$, i.e.,

$$V^* = -U\left(V^+(U^T)^{-1}\right). \quad (13)$$

Therefore, after the column replacement as in Eq. (11), matrices of one-quasiparticle states, $U^{(\mu)}_{\nu'\nu}$, become singular and have null spaces of dimensions $D=1$. Consequently, the corresponding matrices $1 - \rho^{(\mu)} = U^{(\mu)}U^{(\mu)+}$, cf. Eqs. (4) and (5), have exactly *one* eigenvalue equal to zero. Hence, the occupation numbers (eigenvalues of the density matrices $\rho^{(\mu)}$) of *one* of the single-particle states are in each case equal to 1. This fact is at the origin of the name "blocked states" attributed to one-quasiparticle states (12). These states contain fully occupied single-particle states that do not contribute to pairing correlations.

We can continue by building two-quasiparticle states

$$|\Phi\rangle^{(\mu\mu')}_{\text{even}} = \mathcal{N}\alpha^+_\mu \alpha^+_{\mu'} \exp\left(-\tfrac{1}{2}\sum_{\mu\nu} Z^+_{\mu\nu} a^+_\mu a^+_\nu\right)|0\rangle, \quad (14)$$

which are manifestly particle-number parity even, $\pi_N = 1$. They correspond to two columns in $\varphi$ replaced by two columns of $\chi$, and the corresponding matrices $U^{(\mu\mu')}$ have null spaces of dimensionality $D=0$ or $D=2$. We now have a very definite prescription for telling which Bogoliubov transformations correspond to even and which to odd states, i.e., $\pi_N = (-1)^D$, where $D$ is the dimensionality of the null space of matrix $U$.

The main conclusion of this section is that vacuum states of given $\pi_N$ are obtained by making appropriate choices of occupied quasiparticle states. In particular, one should begin by selecting one even state (10) represented by a non-singular matrix $U$, which one can call a reference state, and than proceed by building on it one-, two-, or many-quasiparticle excitations. Note that constructing odd states is best realized by first building an even reference state and then making one-quasiparticle excitations thereof. This is best done by blocking specific quasiparticles, i.e., replacing columns of matrices as in Eq. (11). After the self-consistent procedure is converged for each blocked state, one may select the lowest one as the ground state of an odd system, and consider the higher ones as good approximations of the excited odd states. It is obvious that self-consistent polarization effects exerted by blocked states, which will be taken into account by iterating Eq. (7), may render reference states of every blocked configuration to be different from one another.

Note also that in the above analysis we did not talk about the average particle numbers, which can be even, odd, or fractional, depending on the value of the Fermi energy $\lambda$ in Eq. (6). Thus one can, in principle, consider odd states with even average particle numbers, or even states with odd average particle numbers. The latter ones provide especially useful reference states for building one-particle excitations on top of them, because they require the smallest readjustment of the Fermi energy between the reference state and a one-quasiparticle excitation.

### C. Signature symmetry

In this section we apply methods of occupying quasiparticle states, outlined in Sec. II A, to a physical situation where the system has a conserved signature symmetry, which can be used within the cranking approximation [17, 18, 21]. The signature operation [22] is a rotation by $\pi$ around one direction in space, which is conventionally called the $x$-axis:

$$\hat{R}_x = \exp(-i\pi\hat{J}_x), \quad (15)$$

where $\hat{J}_x$ denotes the total angular-momentum operator along the $x$ axis. The signature operator is manifestly unitary, $\hat{R}^+_x \hat{R}_x = 1$. Since a rotation by $2\pi$ reverses the phase of fermion wave functions, the square of the signature operator gives the particle-number parity, $\hat{R}^2_x = \pi_N$. Therefore, $\hat{R}_x$ is hermitian and antihermitian in even and odd spaces, respectively. In particular, in the single-particle space, the signature is a unitary antihermitian operator.

Since the signature $\hat{R}_x$ and time-reversal $\hat{T}$ operators commute,

$$\hat{T}^+ \hat{R}_x \hat{T} = \hat{R}_x, \qquad (16)$$

signature is a time-even operator. Therefore, for non-rotating systems (i.e., without time-odd fields), signature is equivalent to the time-reversal symmetry $\hat{T}$, but it is more convenient to use, because $\hat{R}_x$ is a linear and not an antilinear operator [23]. For states with nonzero angular momentum, where $\hat{T}$ is internally broken, $\hat{R}_x$ is often still preserved [23, 24, 25]. The single-particle (and one-quasiparticle) states may then be classified according to the signature exponent quantum number $\alpha$ [26]:

$$\hat{R}_x |\alpha k\rangle = e^{-i\pi\alpha} |\alpha k\rangle, \qquad (17)$$

where $\alpha$ takes the values of $\pm 1/2$. For conserved signature, the HFB mean fields $h$ and $\Delta$ commute and anti-commute with $\hat{R}_x$, respectively, and the HFB equations (7) can be written in a good signature basis:

$$\mathcal{H}_\alpha |\alpha\mu\rangle = E_{\alpha\mu} |\alpha\mu\rangle, \qquad (18)$$

where the HFB Hamiltonian matrix (6) in one signature reads

$$\mathcal{H}_\alpha = \begin{pmatrix} (h-\lambda)_{\alpha k, \alpha k'} & \Delta_{\alpha k, -\alpha k'} \\ \Delta^\dagger_{-\alpha k, \alpha k'} & (-h+\lambda)_{-\alpha k, -\alpha k'} \end{pmatrix} \qquad (19)$$

and the two-component quasiparticle wave function is

$$|\alpha\mu\rangle = \begin{pmatrix} U^{\alpha\mu} \\ V^{\alpha\mu} \end{pmatrix}. \qquad (20)$$

A quasiparticle state with good signature is a linear combination of states in time-reversed orbits [23]:

$$|1/2, k\rangle = \tfrac{1}{\sqrt{2}} \left[ -|k\Omega_k\rangle + \pi(-1)^{\Omega_k - 1/2} \hat{T}|k\Omega_k\rangle \right] \quad (21)$$

$$|-1/2, k\rangle = \tfrac{1}{\sqrt{2}} \left[ \hat{T}|k\Omega_k\rangle + \pi(-1)^{\Omega_k - 1/2} |k\Omega_k\rangle \right], \quad (22)$$

where $\Omega$ is the eigenvalue of the $z$ component of the single-particle angular momentum, $\hat{J}_z$, and we have adopted the phase convention according to which $\hat{T}|\pi j\Omega\rangle = \pi(-1)^{j+\Omega}|\pi j, -\Omega\rangle$ [27, 28], where $\pi = \pm 1$ is the parity quantum number. If the Kramers degeneracy is present, the description in terms of Kramers doublets and signature doublets is equivalent. It is for polarized systems having time-odd mean fields that the use of the signature symmetry is superior.

The HFB equation (18) has the (quasiparticle-quasihole) symmetry (7): for each state $|\alpha\mu\rangle$ of a given signature, there exists a conjugate state of opposite signature $|-\alpha\tilde{\mu}\rangle$, opposite energy:

$$E_{-\alpha\tilde{\mu}} = -E_{\alpha\mu}, \qquad (23)$$

and the quasiparticle wave function given by

$$|-\alpha\tilde{\mu}\rangle = \begin{pmatrix} V^{\alpha\mu*} \\ U^{\alpha\mu*} \end{pmatrix}. \qquad (24)$$

By this symmetry, one needs to solve the HFB equation (18) only for one signature, obtaining positive and negative quasiparticle energies $E_{\alpha\mu}$. Therefore, the entire set of negative quasiparticle energies is composed of two groups: (i) the negative ones $E_{\alpha\mu}$ obtained directly from the HFB equation solved for signature $\alpha$, and (ii) the inverted positive ones (23), which correspond to states of signature $-\alpha$.

The zero-quasiparticle HFB reference state (10), representing the lowest configuration for a system with even number of fermions, corresponds to a filled sea of Bogoliubov quasiparticles with negative energies (Fig. 1(a)). In a one-quasiparticle state, representing a state in an odd nucleus, a positive-energy state is occupied and its conjugated partner is empty (Fig. 1(b)).

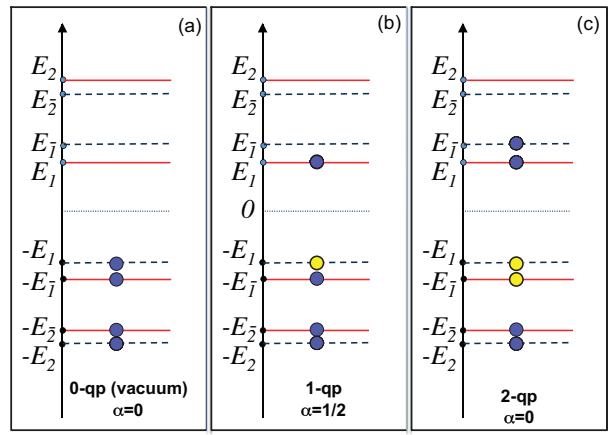

FIG. 1: (Color online) Quasiparticle content of three configurations: (a) vacuum; (b) the lowest one-quasi-particle state with $\alpha=1/2$, accessible via 2FLA; (c) the lowest two-quasi-particle state with $\alpha=0$, not accessible via 2FLA.

The exchange of the eigenvectors $(U^\mu, V^\mu)$ and $(V^{\mu*}, U^{\mu*})$, that have opposite signatures, corresponds to the exchange of columns in the $\varphi$ and $\chi$ matrices discussed in Sec. II A and reverses the particle-number parity $\pi_N$. The density matrix and a pairing tensor of a one-quasiparticle state (12) can be obtained from Eqs. (11) [29, 30]:

$$\rho^{\alpha\mu}_{\alpha k, \alpha l} = \rho^0_{\alpha k, \alpha l} - V^{\alpha\mu}_{-\alpha l} V^{\alpha\mu*}_{-\alpha k} + U^{\alpha\mu*}_{-\alpha l} U^{\alpha\mu}_{-\alpha k}, \quad (25a)$$

$$\kappa^{\alpha\mu}_{\alpha k, -\alpha l} = \kappa^0_{\alpha k, -\alpha l} - U^{\alpha\mu}_{\alpha k} V^{\alpha\mu*}_{-\alpha l} + V^{\alpha\mu*}_{-\alpha k} U^{\alpha\mu}_{-\alpha l}, \quad (25b)$$

where $\rho^0$ and $\kappa^0$ correspond to the reference state (10).

## III. TWO FERMI LEVEL APPROACH

The main idea behind the 2FLA [7, 10] is to force a nonzero spin polarization in the system by finding the ground-state condensate in the presence of an external field that favors one spin over the other. In the language of signature, "spin-up" corresponds to $\alpha = 1/2$ while "spin-down" corresponds to $\alpha = -1/2$. The polarization is achieved by adding to the Hamiltonian the single-particle field

$$h_s = i\lambda_s \hat{R}_x, \quad (26)$$

i.e., constructing the Routhian (9), $h' = h - h_s$. The imaginary unit must be put in the definition of $h_s$, because in the single-particle space $\hat{R}_x$ is antihermitian. Consequently, $h_s$ is time-odd, cf. Eq. (16). Since $\hat{R}_x$ and the quasiparticle Hamiltonian commute, adding $h_s$ represents a non-collective cranking; hence, the quasiparticle routhians must be linear in $\lambda_s$.

The field (26) will raise the Fermi energy $\lambda_{1/2}$ of the subsystem having signature $\alpha = +1/2$ by an amount $\lambda_s$ and lower the Fermi energy $\lambda_{-1/2}$ of the $\alpha = -1/2$ subsystem by the same amount. The relations between chemical potentials in 2FLA read:

$$\lambda_\alpha = \lambda + 2\alpha\lambda_s, \quad (27)$$

where

$$\lambda = \tfrac{1}{2}\left(\lambda_{1/2} + \lambda_{-1/2}\right), \quad \lambda_s = \tfrac{1}{2}\left(\lambda_{1/2} - \lambda_{-1/2}\right). \quad (28)$$

The HFB Routhian matrix of 2FLA can be written as [7]

$$\mathcal{H}_s = \mathcal{H} - 2\alpha\lambda_s \mathcal{I}, \quad (29)$$

where $\mathcal{H}$ is the matrix (19) corresponding to $\lambda_s = 0$ and $\mathcal{I}$ is the unit matrix.

Since the added term is proportional to the unit matrix, its only effect is to shift the HFB eigenvalues up or down,

$$E^s_{\alpha\mu} = E_{\alpha\mu} - 2\alpha\lambda_s, \quad (30)$$

where $E_{\alpha\mu}$ are the eigenvalues of $\mathcal{H}$. Therefore, when plotted as a function of $\lambda_s$, the energies $E^s_{\alpha\mu}$ are straight lines with slopes $2\alpha = \pm 1$, as schematically depicted in Fig. 2. The Hamiltonian (19) usually represents an unpolarized system, in which case the quasiparticle energies $E_{\alpha\mu}$ are degenerate (Fig. 2, top). If $\mathcal{H}$ has time-odd fields (due, e.g., to an external magnetic field or nonzero angular velocity), this Kramers degeneracy is lifted (Fig. 2, bottom).

### A. Interpretation of polarizing field $h_s$

The external field (26) has a particularly simple interpretation for spin systems that are spherically symmetric in space (e.g. in a spherically-symmetric trap). Since in this case we can neglect the orbital part $\hat{L}$ of the angular momentum in $\hat{J} = \hat{L} + \hat{s}$, the polarizing field can be written as

$$h_s = i\lambda_s \exp(-i\pi \hat{s}_x) = 2\lambda_s \hat{s}_x, \quad (31)$$

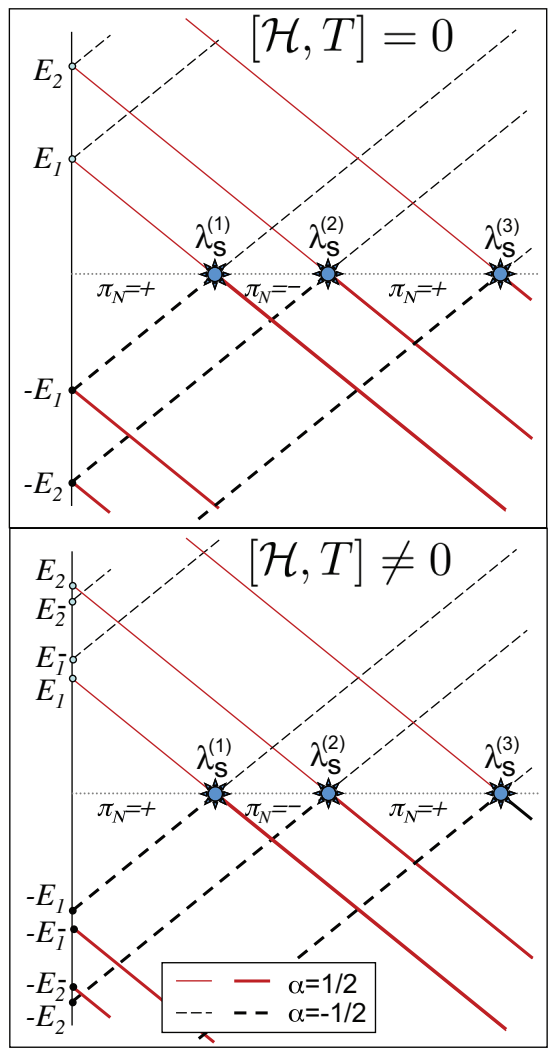

FIG. 2: (Color online) One-quasiparticle levels of both signatures ($\alpha=1/2$: solid line; $\alpha=-1/2$: dotted line) as functions of $\lambda_s$ for a HFB Hamiltonian (19) without (top) and with (bottom) time-odd fields. In the latter case, the Kramers degeneracy at $\lambda_s=0$ is lifted, i.e., states $|\alpha\mu\rangle$ and $\hat{T}|\alpha\tilde{\mu}\rangle$ have different energies. The negative energy levels occupied in the quasiparticle vacuum are marked by thick lines. At the points $\lambda_s^{(n)}$ marked by stars, the quasiparticle level $E_n$ with $\alpha=1/2$ becomes occupied. If this level is not degenerate, or its degeneracy is an odd integer, the particle-number parity $\pi_N$ of the vacuum changes as indicated.

and 2FLA is equivalent to the (one-dimensional) rotational cranking approach [17, 18], in which the total energy of the system at a fixed value of the total angular momentum $J_x = \langle \hat{J}_x \rangle$ is obtained by adding to the Hamiltonian the one-body field $-\omega \hat{J}_x$ (rotational cranking term). Therefore, for spin systems, the angular velocity $\omega$ is simply equal to $2\lambda_s$.

A different situation occurs in atomic nuclei, where the spin degeneracy is lifted by the spin-orbit interaction. Then, projection of the single-particle angular mo-mentum $\Omega_x$ is a good quantum number. Consequently, the eigenvalues of the signature operator (15) are simply $e^{-i\pi\Omega_x}$, i.e., $\alpha$=1/2 for $\Omega_x = 1/2, -3/2, 5/2, \cdots$ and $\alpha$=-1/2 for $\Omega_x = -1/2, 3/2, -5/2, \cdots$. In this situation, there is no difference between signature and spin projection, and odd states can indeed be obtained within the 2FLA. However, in this case, the angular momentum polarization (or spin polarization in a deformed trap rotating along the symmetry axis) is best modeled by the cranking approximation, where $\hat{J}_x$, rather than signature $\hat{R}_x$, is used to build Routhian (9). This is so, because for systems such as nuclei that are governed by $j-j$ coupling, the polarizing field $2\lambda_s\alpha$ does not distinguish between individual single-particle alignments, i.e., between states with large $\Omega_x$ values that predominantly contribute to the angular momentum alignment and low $\Omega_x$ states that weakly respond to rotation. In this respect, the standard non-collective cranking approach has a certain advantage.

Within the non-collective cranking approach, the single-particle cranking term becomes $-\omega\Omega_x$ and the single-quasiparticle energies are linear in $\omega$ with slopes distinguishing between angular momentum projections. The corresponding quasiparticle vacua represent "optimal" states with maximally aligned angular momentum (maximum polarization). The minimization of the total energy for intermediate values of angular momentum can be done by considering the lowest particle-hole excitations across the Fermi surface. These states are not accessible within the 2FLA, and they must be obtained by explicitly blocking quasiparticle states.

### B. Particle-number parity of the HFB vacuum

Let us first analyze the 2FLA in the case when levels depicted in Fig. 2 are not degenerate. At low values of $\lambda_s$, the quasiparticle vacuum corresponds to a system with even number of fermions. It is seen that at $\lambda_s=E_{1/2,1}$ the down-sloping lowest quasiparticle with $\alpha$=1/2 crosses zero and becomes negative. Beyond that point, the HFB vacuum has one quasiparticle state occupied, as in the middle panel of Fig. 1. Here, the particle number parity $\pi_N$ changes from even (+1) to odd (-1), as discussed in Sec. II B and also in the context of nuclear rotations in Refs. [17, 31]. This can also be derived in a straightforward and explicit way by calculating the expectation value of the number-parity operator $e^{i\pi\hat{n}}$ in the HFB vacuum. For each subspace in the canonical representation, the number operator can be expressed $\pi_{N_\alpha}=e^{i\pi(\hat{n}_1+\hat{n}_2)} = (1-2\hat{n}_1)(1-2\hat{n}_2)$, where 1,2 label the quasiparticle transformations for the negative eigenvalues. The ground-state expectation values of $\hat{n}_1$, $\hat{n}_2$, and $\hat{n}_1\hat{n}_2$ are then evaluated in the usual way by expanding in the quasiparticle basis, normal ordering, and extracting the zero-quasiparticle term.

By using the modified density matrices (25) in the HFB equations and in the particle number equation for $\lambda$,

$$N = \text{Tr}(\hat{\rho}), \quad (32)$$

one formally recovers the standard blocked HFB equations for the lowest 1-qp ($\alpha$=1/2) state. At still higher values of $\lambda_s$, the particle-number parity changes again at $\lambda_s=E_{1/2,2}$ when the second lowest quasiparticle with $\alpha$=1/2 crosses zero. The associated two-quasiparticle configuration in a nucleus with $N = N_0 + 2$ has the signature index $\alpha$=1. It is immediately seen that the lowest two-quasiparticle $\alpha$=0 configuration of Fig. 1(c), associated with the so-called S-band in rotating nuclei, cannot be reached within the standard 2FLA.

Following the discussion in Sec. III A, it is worth noting that the choice of angular momentum quantization implies a different character of the angular alignment associated with the change in the quasiparticle vacuum. In the case of $z$-quantization and the absence of time-odd fields in the HFB Hamiltonian, 2FLA treatment of systems with odd particle number is equivalent to the so-called uniform filling approximation, in which a blocked nucleon is put with equal probability in each of the degenerate magnetic substates [32, 33]. It is only in the regime of non-collective rotation in which the angular momentum is quantized along the $x$-axis that the dynamics of angular momentum alignment can be properly treated and the full alignment can be reached.

In cases when in the unpolarized system the Kramers degeneracy is the only one, the parity $\pi_N$ of states obtained by occupying the lowest $E^s_{\alpha\mu} < 0$ quasiparticles do change at points where $E^s_{\alpha\mu} = 0$, as indicated in Fig. 2. However, if apart from the Kramers degeneracy there is an additional two-fold, four-fold, etc. degeneracy of quasiparticle levels, all such states will be $\pi_N$-even. Therefore, in such situations, the 2FLA would fail to produce odd-$N$ systems as HFB ground states.

For spherical spin systems, each level in Fig. 2 is $(2\ell+1)$-degenerate, which is an odd number, and the number parity does change at points $E^s_{\alpha\mu} = 0$. However, the corresponding HFB vacuum represents a $(2\ell+1)$-quasiparticle excitation and not the one-quasiparticle excitation of Eq. (12). This is so even if the average Fermi energy $\lambda$ is adjusted to have (on average) only one particle more than that of the unpolarized system [7, 10]. Such a situation corresponds to the so-called filling approximation of orbitally degenerate quasiparticle states. Needless to say, the orbital filling approximation completely neglects possible space-polarization effects that must, in principle, occur for true one-quasiparticle states.

If the spin system has an axial symmetry in space (e.g., it is in an external axially symmetric trap), projection of the angular momentum on the symmetry axis $\Lambda$ is a good quantum number. Moreover, the single-particle and quasiparticle states are then degenerate with respect to the sign of $\Lambda$. Here, each level in Fig. 2, except for $\Lambda$=0, is doubly degenerate. Therefore, none of the $\Lambda$>0 states obtained by occupying the $E^s_{\alpha\mu} < 0$ quasiparticles has $\pi_N$=−1, that is, odd particle number. In this case, states with $\pi_N=-1$ cannot be obtained within 2FLA, and explicit treatment within the blocking approximation, described in Sec. II C, is mandatory.

## IV. ATOMIC AND NUCLEAR HFB CALCULATIONS

The first numerical example of 2FLA deals with a two-component polarized atomic condensate in a deformed harmonic trap using the superfluid local density approximation [10, 34]. The system is described by a local energy density

$$\mathcal{E}(\boldsymbol{r}) = \alpha_u \frac{\tau(\boldsymbol{r})}{2} + \beta_u \frac{3(3\pi^2)^{2/3}\rho^{5/3}(\boldsymbol{r})}{10} + g_{eff}(\boldsymbol{r})|\kappa(\boldsymbol{r})|^2, \quad (33)$$

where the local densities $\rho(\boldsymbol{r})$ (particle density), $\tau(\boldsymbol{r})$ (kinetic energy density), and $\kappa(\boldsymbol{r})$ (pairing tensor) are constructed from the quasiparticle HFB wave functions. The parameters $\alpha_u$, $\beta_u$, and the effective pairing strength $g_{eff}(\boldsymbol{r})$ have been taken according to Ref. [10]. We assume that the external trapping potential can be described by an axially deformed harmonic oscillator with frequencies [17]

$$\omega_\perp^2(\delta) = \omega_0^2(\delta)\left(1 + \frac{2}{3}\delta\right), \; \omega_x^2(\delta) = \omega_0^2(\delta)\left(1 - \frac{4}{3}\delta\right), \quad (34)$$

with

$$\omega_0(\delta) = \tilde{\omega}_0\left(1 + \frac{2}{3}\delta^2\right). \quad (35)$$

As in Ref. [10], we put $\hbar\tilde{\omega}_0=1$.

The calculations were carried out for systems with $N=30$ and 31 fermions in a spherical ($\delta=0$) and deformed ($\delta=0.2$) trap. The HFB equations were solved by using the recently developed axial DFT solver HFB-AX [35]. The results are displayed in Fig. 3.

In the spherical case, Fig. 3 (top), the spin degeneracy is lifted by the polarizing field $h_s$. However, as discussed in Sec. III B, the orbital $(2\ell+1)$-fold degeneracy is present. Consequently, after the crossing point, the vacuum becomes a $(2\ell_1+1)$-quasiparticle state, where $\ell_1$ is the orbital angular momentum of the lowest quasiparticle level. For $N=30$, the lowest quasiparticle excitation is a $p$ state, i.e., above the crossing point the local HFB vacuum becomes a three-quasipartiicle state. At the crossing point, the self-consistent mean-field changes abruptly. In particular, the chemical potential moves up as the number of particles increases by one, and the pairing gap decreases due to blocking. This produces a sharp discontinuity around the crossing point, which can be seen in all three cases presented in Fig. 3.

The middle portion of Fig. 3 illustrates the deformed case. Here, quasiparticle states are labeled using the angular momentum projection quantum number $\Lambda$ onto the symmetry axis of the trapping potential ($x$-axis). Because of the Kramers degeneracy, levels with $\pm\Lambda$ are degenerate. That is, except for $\Lambda=0$, each quasiparticle state is two-fold degenerate. In the case presented in Fig. 3, the two lowest levels have $\Lambda=0$, hence they are associated with one-quasiparticle excitations. The third state has $|\Lambda|=1$, and its crossing does not change the particle-number parity of the HFB vacuum.

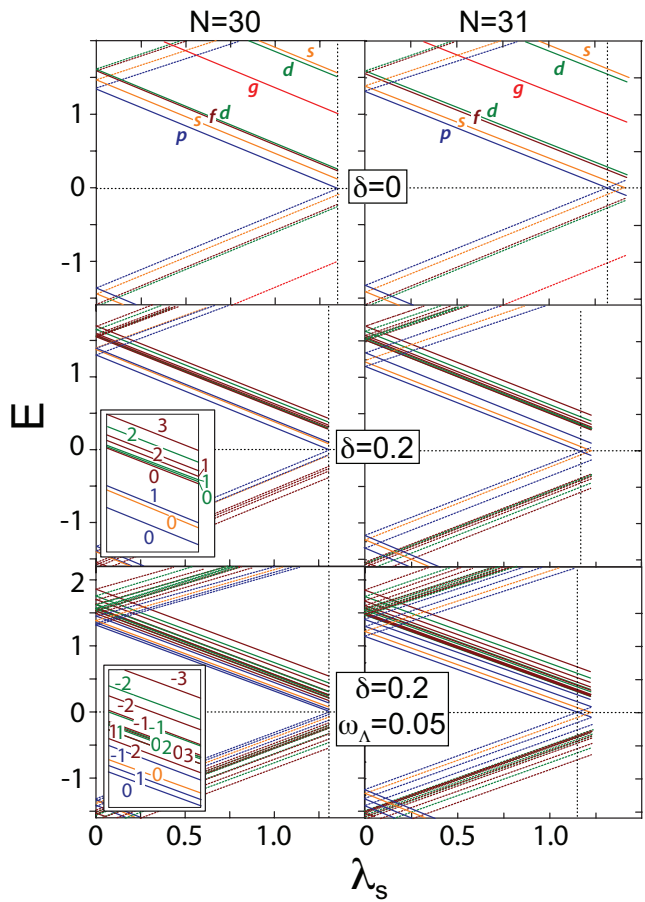

FIG. 3: (Color online) Quasiparticle levels ($\alpha=1/2$: solid line; $\alpha=-1/2$: dotted line) as functions of $\lambda_s$ for a two-component polarized atomic condensate in a deformed harmonic trap using the superfluid local density approximation [10, 34]. The calculations were carried out for $N=30$ (left) and $N=31$ (right). The top panels correspond to the spherical case ($\delta=0$). Here the quasiparticle levels are labeled with the usual spectroscopic designation of orbital angular momentum $\ell$. Each line represents a set of $(2\ell+1)$-fold degenerate states. In the deformed case ($\delta=0.2$; middle panels) the levels with the opposite values of $\Lambda>0$ are two-fold degenerate. The corresponding values of $|\Lambda|$ ere shown in the inset. The Kramers degeneracy can be lifted by adding a cranking term, $-\omega_\Lambda\hat{\ell}_x$ to the Hamiltonian (bottom panels, $-\omega_\Lambda=0.05$). Here, every level is labeled by the $\Lambda$ quantum number, see the inset. In order to make the plot less busy, the quasiparticle levels originating from very excited spherical $g$, $d$, and $s$ shells are not shown in the two lower panels. See text for details.

The two-fold Kramers degeneracy can be removed by adding an external orbital-polarizing field,

$$\hat{h}_\ell = -\omega_\Lambda \hat{\ell}_x, \quad (36)$$

where $\omega_\Lambda$ is the cranking frequency for the orbital motion. Indeed, in the absence of the spin-orbit coupling the spin and the orbital angular momentum may rotate with different angular velocities. In the presence of the field (36), each level is shifted by $-\omega_\Lambda \Lambda$, i.e., the energy splitting of the Kramers doublet becomes $2\omega_\Lambda |\Lambda|$. An illustrative example of such situation is displayed in the bottom panels of Fig. 3. Here, each level corresponds to a one-quasiparticle excitation. We confirmed numerically that the result of calculations for $N=31$ by explicitly blocking the lowest level are here equivalent to those with 2FLA carried out above the crossing. It is seen in Fig. 3, however, that because of high density of quasiparticle levels, the self-consistent calculations in 2FLA are difficult due to many consecutive crossings that make it extremely difficult to keep track of the fixed configuration. We note that the order of the quasiparticle levels in $N=30$ and $N=31$ systems is affected by the variation of the mean field due to the crossing.

In order to illustrate 2FLA in the nuclear case, we carried out nuclear DFT calculations using the Skyrme energy density functional SLy4 [36] in the p-h channel, augmented by the "mixed-pairing" [37] density-dependent delta functional in the p-p channel. The details pertaining to the numerical details, e.g. the pairing space employed, can be found in Ref. [35]. As a representative example, we took the pair of deformed nuclei $^{166}$Er ($N=98$) and $^{167}$Er ($N=99$). The pairing strength $V_0=-320\,\text{MeV}\,\text{fm}^3$ was slightly enlarged to prevent pairing from collapsing in the $N=99$ system. The resulting neutron pairing gaps, $\Delta_n=1.2\,\text{MeV}$ and $0.77\,\text{MeV}$ in $^{166}$Er and $^{167}$Er, respectively, are reasonably close to the experimental values of $1.02\,\text{MeV}$ and $0.62\,\text{MeV}$.

Figure 4 displays the quasiparticle spectrum for $^{166}$Er (top) and $^{167}$Er (bottom). At the value of $\lambda_s$ indicated by a star symbol, a transition from a zero-quasiparticle vacuum corresponding to $N=98$ to a one-quasiparticle vacuum associated with $N=99$ takes place. As in the atomic case, the mean-field changes abruptly at the crossing point. Actually, since the quasiparticle spectrum changes when going from $^{166}$Er to $^{167}$Er (both in terms of excitation energy and ordering of levels), the crossing point is shifted towards the lower values of $\lambda_s$. As checked numerically, the result of calculations for $^{167}$Er by explicitly blocking the level "a" are equivalent to those with 2FLA carried out above the crossing point.

## V. CONCLUSIONS

In this study, we analyzed various approaches to polarized Fermi systems within the DFT. The main conclusions can be summarized as follows:

- By analogy with rotating nuclei, we showed that introducing two chemical potentials for different

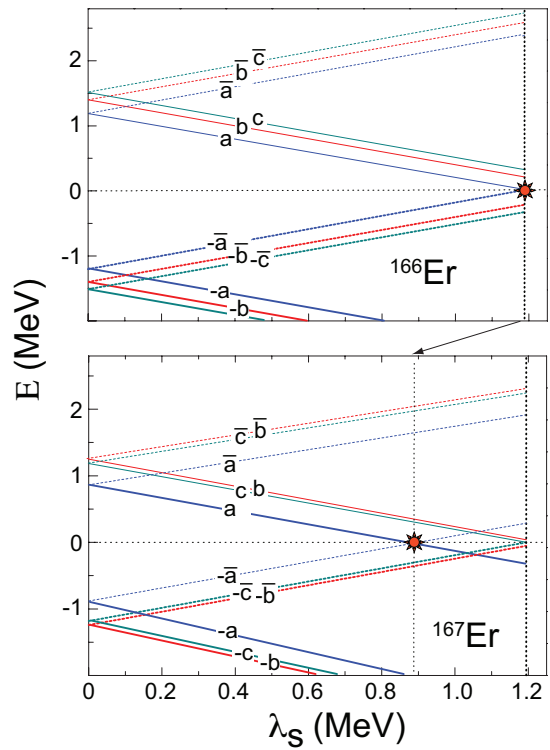

FIG. 4: (Color online) One-quasiparticle levels of both signatures ($\alpha=1/2$: solid line; $\alpha=-1/2$: dotted line) as functions of $\lambda_s$ for $^{166}$Er (top) and $^{167}$Er (bottom). The levels occupied in the vacuum configuration are drawn by thick lines. The calculations were carried out with a SLy4 Skyrme functional and mixed pairing. See text for details.

superfluid components is equivalent to applying a one-body, time-odd field. This field can be used to change the particle-number parity of the underlying quasiparticle vacuum.

- Since the external one-body field commutes with the HFB Hamiltonian, 2FLA is equivalent to non-collective cranking, a technique that is often used to select a vacuum configuration of interest.

- For systems, in which no additional degeneracy is present beyond the Kramers degeneracy, the 2FLA is equivalent to one-dimensional, non-collective rotational cranking. Different choices of the angular momentum quantization axis give rise to different blocking procedures and different polarization schemes. By increasing the asymmetry $\lambda_s$, one is alternating between even and odd systems while gradually increasing the signature polarization. In the absence of the spin-orbit coupling, the signature quantum number can be replaced by spin projection.

- The generalization of 2FLA to the case of an arbitrary blocked state is not obvious, although one can introduce polarizing fields that would single out the state of interest. For instance, one can introduce different cranking frequencies for spin and orbital motion, and this removes the orbital degeneracy of quasiparticle states. However, the standard way of blocking a given level in a fixed Hamiltonian submatrix should work as well, and it is easy to generalize to an arbitrary state, even if the majority of self-consistent quantum numbers are gone.

- For systems, in which an additional $k$-fold degeneracy ($k$-even) is present beyond the Kramers degeneracy, the 2FLA is unable to produce odd-particle-number states. Moreover, for an $k$-fold degeneracy with odd $k$, which is the case, e.g., for spherical systems, the 2FLA gives odd-particle-number states that correspond to $k$-quasiparticle and not to one-quasiparticle excitations.

- The situation encountered in Fig. 2 is similar to the level crossing discussed in the context of high-spin physics, where the lowest quasiparticle Routhian becomes negative as a function of rotational frequency. In such situations, one needs to preserve a number of quasiparticles in each signature block to conserve $\pi_N$ in the HFB vacuum [29, 31].

- Irrespective of the degeneracies present in the system, certain quasiparticle configurations cannot be approached through 2FLA.

In summary, the 2FLA provides a unifying methodology to treat a number of different kinds of condensates, including those of odd-particle systems, as ground states of some HFB Hamiltonian. However, not all quasiparticle states are easily accessible this way and problems arise if quasiparticle levels show degeneracies beyond the Kramers doubling. Moreover, the examples shown in Figs. 3 and 4 indicate that the variations of the self-consistent mean field associated with the configuration change driven by the polarizing field can be severe, including the change of ordering of the lowest quasiparticle excitations. This, together with related numerical instabilities, indicates that traditional methods of blocking are likely to be preferred for treating a large space of quasiparticle configurations and in spectroscopic-quality calculations for well-defined one-quasiparticle states.

Useful discussions with Aurel Bulgac, Michael Forbes, and Mario Stoitsov are gratefully acknowledged. The UNEDF SciDAC Collaboration is supported by the U.S. Department of Energy under grant No. DE-FC02-07ER41457. This work was also supported by the U.S. Department of Energy under Contract Nos. DE-FG02-96ER40963 (University of Tennessee), DE-AC05-00OR22725 with UT-Battelle, LLC (Oak Ridge National Laboratory), DE-FG05-87ER40361 (Joint Institute for Heavy Ion Research), and DE-FG02-97ER41014 (University of Washington); by the Polish Ministry of Science under Contract No. N N202 328234 and by the Academy of Finland and University of Jyväskylä within the FIDIPRO programme.